\documentclass[twocolumn,showpacs,preprintnumbers,amsmath,amssymb]{revtex4}
\usepackage{color}
\usepackage{graphicx}% Include figure files
\usepackage{dcolumn}% Align table columns on decimal point
\usepackage{bm}% bold math
\begin{document}

\preprint{APS/123-QED}

\title{Novel Polaron State for Single Impurity in a Bosonic Mott Insulator}% Force line breaks with \\

\author{Yasuyuki Kato$^1$}

\author{K. A. Al-Hassanieh$^1$}

\author{A. E. Feiguin$^2$}

\author{Eddy Timmermans$^1$}

\author{C. D. Batista$^1$}

\affiliation{$^1$Theoretical Division, Los Alamos National Laboratory, Los Alamos, New Mexico 87545, USA}

\affiliation{ $^2$ Department of Physics and Astronomy, University of Wyoming, Laramie, WY 82071, USA}

\date{\today}% It is always \today, today,

\begin{abstract}
We show that a single impurity embedded in a cold atom bosonic Mott insulator 
leads to a novel polaron that exhibits correlated motion with an effective mass and a linear size that nearly 
diverge at critical value of the on-site impurity-boson interaction strength.  
Cold atom technology can tune the polaron's properties and break up the composite particle into a
deconfined impurity-hole and boson particle state at finite, controllable polaron momentum.
\end{abstract}

\pacs{Valid PACS appear here}% PACS, the Physics and Astronomy
\maketitle

The exploration of and unprecedented control over quantum many-body systems have become central 
driving forces in cold-atom physics.  Optical lattices (the standing wave patterns 
of frequency-stable, reflected laser beams that are experienced by ultra-cold atoms as 
periodic potentials \cite{greiner2008}) have unlocked strongly-correlated lattice physics to cold
atom simulation.  The Bose-Hubbard Hamiltonian \cite{jaksch1998} has
successfully and quantitatively modeled boson atom dynamics
in such optical lattices.  Studies of the boson superfluid to Mott
insulator (MI) phase transition predicted within the Bose-Hubbard
model \cite{fisher1989} highlight the unusual cold atom access and control:
experiments observed the transition \cite{greiner2002}, revivals of inter-site
superfluid coherence \cite{greiner2002b} and different integer filling-number
islands in the MI phase as seen spectroscopically \cite{campbell2006}
and, more recently, by direct imaging \cite{bakr2009,sherson2010,bakr2010,hung2010}.  In this Letter, we
show that by combining the control over optical lattice parameters
(varying the barrier height), over the inter-particle interactions
(varying an external, homogeneous magnetic field in a Feshbach
resonance \cite{inouye1998,bloch2008}), and over particle-species (creating mixtures of
distinguishable kinds of atoms), cold atom experiments can
realize a novel \footnote{The word 'novel' here refers not only to the cold atom environment
but also to the properties of the polaron. } and controllable polaron in the MI phase.

The polaron state is induced by an impurity atom that experiences the same (or similar) optical lattice 
potential as the MI bosons from which it is distinguishable. 
The polaron consists of the 
impurity and a boson that is promoted to the next Hubbard band by a strong impurity-boson 
repulsion.  If the impurity-boson interaction is Feshbach tuned to be nearly as repulsive as the boson-boson 
interaction, the excited boson remains pinned to the impurity site.  The boson-impurity pair propagates 
through the lattice with finite total momentum ${\bf K}$.
The linear polaron size, $\lambda_{\bf K}$, or average distance between the impurity and the excited boson 
sensitively depends on ${\bf K}$, and increases with increasing impurity-boson repulsive interaction $U_{IB}$
\footnote{The increase in polaron size with increasing $U_{IB}$ contrasts with the size of an impurity atom
in boson superfluid}. Surprisingly, {\it the effective mass of the polaron increases with its size.} Moreover, 
in the strong coupling limit the effective mass diverges at the critical value of $U_{IB}$ above which the 
boson-impurity pair becomes unbound ($\lambda_{\bf 0} \to \infty$). 
Most significantly, optical lattice experiments can create probe and manipulate composite particles with properties
that are unusual in traditional polaronic 
and excitonic systems.

We will model our problem with a bosonic Hubbard Hamiltonian on a hyper-cubic lattice of
dimension $d$:
\begin{eqnarray}
	{\mathcal H} &= & -t_{B} \sum_{\langle {\bf r},{\bf r'} \rangle} b^{\dagger}_{\bf r} b^{\;}_{\bf r'} 
	-t_{I} \sum_{\langle {\bf r},{\bf r'} \rangle} c^{\dagger}_{\bf r} c^{\;}_{{\bf r'}}
	\nonumber\\
	&&
	+ \frac{U_{BB}}{2} \sum_{\bf r} b^\dag_{\bf r}b^\dag_{\bf r}b_{\bf r}b_{\bf r} + U_{IB} \sum_{\bf r} b^\dag_{\bf r}b_{\bf r} c^\dag_{\bf r}c_{\bf r}.
	%&&+ \frac{U_{BB}}{2} \sum_j n^B_{\bf r} (n^B_{\bf r}-1) + U_{IB} \sum_{\bf r} n^B_{\bf r} n^I_{\bf r}.
\end{eqnarray}
The operator $b^{\dagger}_{\bf r}$ ($b^{\;}_{\bf r}$) creates (annihilates) a boson on site ${\bf r}$, while 
$c^{\dagger}_{\bf r}$ ($c^{\;}_{\bf r}$) creates (annihilates) the impurity. 
$\langle {\bf r},{\bf r'} \rangle$ indicates  that ${\bf r}$ and ${\bf r'}$ are nearest-neighbors.
The statistics of $c^{\dagger}_{\bf r}$ and $c^{\;}_{\bf r}$ is irrelevant because we are considering 
a {\it single} impurity problem. 
%The $b$ and $c$ operators commute with each other. 
%The density operators are defined in the usual way: $n^B_{\bf r}=b^{\dagger}_{\bf r} b^{\;}_{\bf r}$ 
%and $n^I_{\bf r}=c^{\dagger}_{\bf r} c^{\;}_{\bf r}$. 

\begin{figure}[!htb]
	 \includegraphics[trim= 0 0 0 0,angle=270,width=7cm,clip]{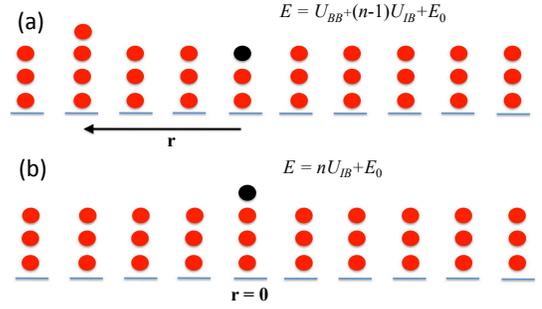}
	 \caption{(color online)Lowest energy eigenstates of a MI with $n$ bosons per site ($n=3$ in the figure) doped
 	with a single impurity (black particle). (a) The ``particle'' and the ``hole'' occupy different positions and
 	the vector ${\bf r}$ is the relative position. (b) The ``particle'' and the ``hole'' occupy the same position (${\bf r}=0$).}
 	\label{lowenstates}
 \end{figure}

Here we will only consider the case of strongly repulsive on-site boson-boson interaction, 
$U_{BB} \gg n|t_{B}|, |t_{I}|$, and integer filling factor 
$\langle b^\dag_{\bf r} b_{\bf r} \rangle = n$ to 
stabilize the MI state.  An increase in the intensity of the optical lattice laser
of wavelength $\lambda$ so that the lattice height $V_{o}$ significantly exceeds the 
recoil energy $\hbar \omega_{R}$ ($= h^{2}/[2m_{B} \lambda^{2}]$), where $m_{B}$ is the
actual boson mass) drives the boson system deep into the MI regime.  This increase
tightens the trapping frequency $\omega_{T}$ of the optical lattice wells, $\omega_{T}=2\omega_{R}
\sqrt{x}$, where $x=V_{0}/\hbar \omega_{R}$.  For fixed boson-boson scattering length
$a_{BB}$, an increase in $x$, $x > 1$, enhances $U_{BB}\approx \hbar \omega_{T}
\sqrt{8\pi} \left(a_{BB}/\lambda\right) x^{1/4}$, and exponentially decreases the hopping
matrix element $t_{BB}$ which we estimate as $t_{B} \approx (3/2) \hbar \omega_{T}
e^{-(\pi/2)^{2}\sqrt{x}}$.  Scattering lengths take the value of a few nm whereas the optical
wavelength is of order of a micron but even for $x\sim 9$, $U_{BB}$ can
exceed $t_{B}$ by an order of magnitude whereas $t_{B}$ can still be of the order of 10 Hz.  Thus,
the MI-regime can be accessed by varying the optical lattice height, which leaves 
a homogeneous magnetic field tuned near a Feshbach resonant value as a control
knob to vary $a_{BB}$ ($\propto U_{BB}$) or the impurity-boson scattering length
$a_{IB}$ ($\propto U_{IB}$) independently.  We consider the regime
$U_{IB} \gg n |t_{B}|, |t_{I}|$.  The eigenstates of ${\mathcal H}$ become highly degenerate 
in the static limit $t_{B}=t_{I}=0$.  The lowest-energy eigenstates, illustrated in Fig.\ref{lowenstates},
can be expressed as
\begin{eqnarray}
	|{\bf R},{\bf r} \rangle &=& \frac{1}{\sqrt{n(n+1)}} c^{\dagger}_{\bf R} b^{\dagger}_{{\bf R}+{\bf r}} b^{\;}_{\bf R}| 0 \rangle
	\;\; {\rm for} \;\; {\bf r \neq 0},
	\label{low1} \\
	|{\bf R},{\bf 0} \rangle &=& c^{\dagger}_{\bf R} | 0 \rangle,
	\label{low0}
\end{eqnarray}
where
$|0 \rangle = \prod_{\bf r} {({n}!)^{-1/2}}b^{\dagger n}_{\bf r} |\emptyset \rangle$, and where
$ |\emptyset \rangle$ denotes the vacuum state.
The states $ |{\bf R},{\bf r} \rangle$ correspond to the impurity
at site ${\bf R}$ creating a hole at the same site and relocating
the removed boson to site ${\bf r}$ (see Fig. \ref{lowenstates}(a)).
%The states $|{\bf R},{\bf r} \rangle$ correspond to the impurity creating a hole on site ${\bf R}$ and promoting 
%the removed boson to the next Hubbard band. 
The size of the resulting polaron is the size of the particle-hole bound state, i.e., the mean value of ${\bf r}$. 
%In particular, the ${\bf r =0}$ states do not contain any distortion of the Mott background due to the presence 
%of the impurity. 

The energy eigenvalues of the states \eqref{low0} are given by
\begin{eqnarray}
	{\mathcal H} |{\bf R},{\bf r} \rangle &=& (U_{BB} + U_{IB} (n-1) + E_0) |{\bf R},{\bf r} \rangle
	\;\; {\rm for} \;\; {\bf r \neq 0}
	\nonumber \\
	{\mathcal H} |{\bf R},{\bf 0} \rangle &=&  (n U_{IB} + E_0) |{\bf R},{\bf 0} \rangle
	\label{lowen}
\end{eqnarray}
where $E_0= N U n (n-1)/2$ represents the ground state energy of the undoped 
MI: ${\mathcal H} |0 \rangle = E_0  |0 \rangle$ 
($N$ is the total number of lattice sites). 
The energy of any other eigenstate is of order $U_{BB}$ or $U_{IB}$ 
higher than the energy of the eigenstates \eqref{lowen}. 
According to perturbation theory, to lowest order in the small parameters 
$t_{\nu}/U_{BB}$ and $t_{\nu}/U_{IB}$ ($\nu=I,B$), we can exclude those high-energy
states and restrict the action of ${\mathcal H}$ to the lowest energy subspace generated 
by the states \eqref{lowen}. In this way we reduce the many-body problem to a two-body 
system described by the low-energy effective Hamiltonian, 
\begin{eqnarray}
{\bar {\mathcal H}} &=& \sum_{\bf K}{\bar {\mathcal H}}_{0{\bf K}}+{\bar {\mathcal H}}_{1{\bf K}}, \nonumber \\
{\bar {\mathcal H}}_{0{\bf K}} &=& -t_B (n+1) \sum_{\langle {\bf r}, {\bf r'} \rangle} |{\bf K}, {\bf r} \rangle 
\langle {\bf K}, {\bf r'}|,
\nonumber \\
{\bar {\mathcal H}}_{1{\bf K}} &=&   \frac{\tau}{\sqrt{2d}} \sum_{\langle {\bf 0}, {\bf r'} \rangle} (|{\bf K}, {\bf 0} \rangle 
\langle {\bf K}, {\bf r'}| + |{\bf K}, {\bf r'} \rangle 
\langle {\bf K}, {\bf 0}|)
\nonumber \\
&&+ U_{\bf K} |{\bf K}, {\boldsymbol 0} \rangle 
\langle {\bf K}, {\bf 0}|,
\end{eqnarray} 
where 
$\tau=  t_B \sqrt{2d} [n+1-\sqrt{n(n+1)}] $,
\begin{equation}
U_{\bf K}= U_{IB} - U_{BB} +\epsilon^I_{\bf K}, \;\;\;
\epsilon^I_{\bf K} =- t_{I} \sum_{\langle {\bf 0}, {\bf r'} \rangle} e^{i{\bf K} \cdot {\bf r'}},
\end{equation}
and
\begin{eqnarray}
|{\bf K},{\bf r} \rangle &=& \frac{1}{\sqrt{Nn(n+1)}} \sum_{\bf R}  e^{i{\bf K} \cdot {\bf R}}c^{\dagger}_{\bf R} b^{\dagger}_{{\bf R}+{\bf r}} b^{\;}_{\bf R}| 0 \rangle
\;\; {\rm for} \;\; {\bf r \neq 0},\nonumber \\
|{\bf K},{\bf 0} \rangle &=& \frac{1}{\sqrt{N}} \sum_{\bf R}  e^{i{\bf K} \cdot {\bf R}} c^{\dagger}_{\bf R} | 0 \rangle = 
c^{\dagger}_{\bf K} | 0 \rangle.
\label{lowk0}
\end{eqnarray}
${\bf K}$ is the center-of-mass momentum of the effective two-particle system. 
Since ${\bf K}$ is a conserved quantity, the 
states with a given ${\bf K}$ generate an invariant subspace of ${\bar {\mathcal H}}$ leading to 
an effective single-particle problem in the relative coordinate ${\bf r}$.
This effective particle moves in a ${\bf K}$ dependent central potential given by
$V({\bf r})=U_{\bf K}\delta_{{\bf r},{\bf 0}}$. 
The hopping amplitude is $(n+1)t_{B}$ except for bonds containing the origin 
where it is equal to $\sqrt{n(n+1)}t_B$.
In an optical lattice, the impurity atom can be accelerated with forces that are invisible to 
the $b$-bosons by using the recently demonstrated species-specific dipole potentials. 
\cite{catani2009,lamporesi2010}

%%%%%%%%%%%%%%%%%%%%%%%%%%%%%

{\it One-Dimensional Case.} Since the potential $V({\bf r})$ can be attractive, we look for
bound-state solutions of ${\bar {\mathcal H}}_{\bf K}$. 
For $d=1$, the exact ground state of ${\bar {\mathcal H}}_{\bf K}$ can be expressed as
\begin{eqnarray}
|\psi^0_{\bf K} \rangle = \alpha_{\bf K} 
\left[
\frac{(n+1)}{\sqrt{n(n+1)}} | {\bf K}, {\bf 0} \rangle + \sum_{\bf r \neq 0} e^{- r/\lambda_{\bf K}} | {\bf K}, {\bf r}\rangle
\right],
\label{wf}
\end{eqnarray}
with $\alpha^2_{\bf K} =   n(1-e^{-1/\lambda_{\bf K}}) / ( 1+n-e^{-1/\lambda_{\bf K}})$, and
\begin{equation}
e^{1/\lambda_{\bf K}} = - \frac{U_{\bf K}}{2t_B(n+1)} 
+ \sqrt{\frac{U^2_{\bf K}}{4t^2_B(n+1)^2}+\frac{(n-1)}{(n+1)}}.
\end{equation}
The corresponding eigenvalue is
\begin{equation}
\epsilon^b_{\bf K} =-2t_B(n+1)\cosh(1/\lambda_{\bf K}).
\end{equation}
%The length ${\lambda_{\bf K}}>0$ is the characteristic size of the polaron with momentum ${\bf K}$. 
This equation has a physical solution (${\lambda_{\bf K}}>0$) for 
$U_{{\bf K}} < U_c$ with $U_c=-2t_B$:
\begin{equation}
 \lambda_{{\bf K}} =\frac{2t_Bn}{U_c - U_{{\bf K}}}.
\end{equation}
Note that the size of the polaron, $\lambda_{\bf 0}$, diverges for $ U_{\bf 0} \rightarrow U_c$. 
The solutions for $U_{{\bf K}} > U_{c}$ are unbound
particle-hole states: the impurity-hole item and the boson that
was promoted to the next Hubbard band propagate independently
while scattering in each other's vicinity.

The attractive potential $ U_{\bf K}$ has to reach a critical value for the stabilization of the bound state 
because the hopping amplitude is smaller for the bonds that include the origin. 
Since we are assuming that $t_I, t_B > 0$, $\epsilon^0_{\bf K}$ has its minimum at ${\bf K=0}$. 
The effective mass, $m_I^*$, of the exciton is given by the equation:
\begin{equation}
\frac{1}{m^*} \equiv \partial^2_{\bf k} \epsilon^b_{\bf K}|_{\bf K=0}
=\frac{|\langle \psi^0_{\bf 0}| {\bf K}, {\bf 0} \rangle|^2 }{m_I}
\label{eq:meffdef}
\end{equation}
where $m_I=(2|t_I|)^{-1}$ is the bare (lattice) mass  of the impurity
in units in which the unit length is given by the lattice constant and $\hbar=1$. 
The identity \eqref{eq:meffdef}
follows from the Hellmann-Feynman theorem and
$\partial_{{\bf K}} \langle \psi^{0}_{{\bf K}} | {\bf K}, {\bf 0} \rangle
|_{{\bf K}=0} = 0$.
%application of the Hellmann-Feynman theorem and the fact that 
%$\partial_{\bf K} \langle \psi^0_{\bf K}| {\bf K}, {\bf 0} \rangle|_{{\bf K}=0}=0$.
By taking the large $\lambda_{\bf 0}$ limit, we obtain the relation between 
$m^*/m_I$ and $\lambda_{{\bf 0}}$ near the critical point:
\begin{equation}
\frac{m^*}{m_I} = \frac{n\lambda_{{\bf 0}}}{n+1}.
\label{rel}
\end{equation}
In the strong coupling regime the effective mass of the MI impurity
polaron is proportional to its size.  This behavior, very different from that of
lattice polarons induced by electrons in condensed matter, is caused by the unusual mode
of transportation: 
%only in the Hilbert subspace in which the displaced boson and the hole 
%mutually annihilate (${\bf r}=0$) can the impurity hop.  
the impurity can hop only when the displaced boson and the hole mutually annihilate ($\bf r$=0).
As a consequence of this correlated motion, the effective mass is proportional
to $|\langle \psi^0_{\bf 0}| {\bf K}, {\bf 0} \rangle|^2$ (see Eq.\eqref{eq:meffdef}).
%%%%%%%%%%%%%%%%%%%%%%%%%%%%%%%%%%%%%%%%%%%%%%%%%
%This is a rather unusual result in comparison with the behavior of lattice polarons induced by electrons in condensed matter
%systems. The crucial point behind Eq.\eqref{rel} is that, to lowest order in $t_{I}$, the impurity can only move to its nearest neighbors
%when the relative coordinate ${\bf r}$ is equal to zero. This is the reason why $m^*/m_I$ turns out to be 
%proportional to $|\langle \psi^0_{\bf 0}| {\bf K}, {\bf 0} \rangle|^2$ [see Eq.\eqref{eq:meffdef}].

\begin{figure}[!htb]
	 \includegraphics[trim= 0 0 0 0,angle=270,width=7.5cm,clip]{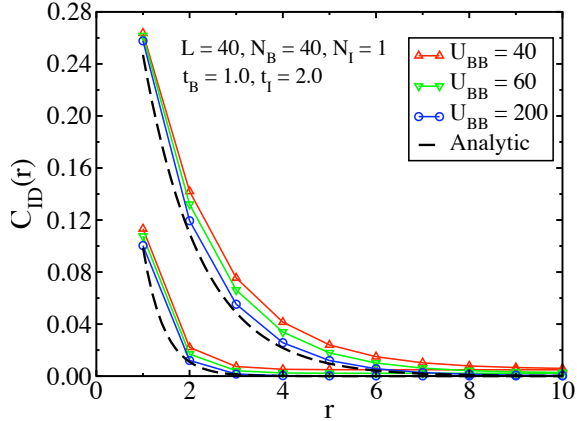}
	 \caption{(color online) Correlation function between the positions of the impurity and the 
site with one additional boson for $n=1$. The dashed line is the analytical result of Eq.\eqref{wf}. The 
full lines were obtained by solving ${\cal H}$ (with a constraint of no more than two particles per site) 
by means of the DMRG method in a chain of $L=40$ sites. 
The upper (lower) curves correspond to
$U_{IB} = U_{BB} + t_B$ ($U_{IB} = U_{BB} - 2t_B$).}
 	\label{dmrg}
 \end{figure}

To test the validity of ${\bar {\mathcal H}}$ we computed the correlation function $C_{ID}(\bf r)$ 
between the positions of the impurity and the site with one additional boson by solving the original 
Hubbard model with the constraint of no more than two bosons per site. 
We used the Density-Matrix Renormalization Group (DMRG)\cite{white1992,white1993}.
Figure \ref{dmrg} shows the comparison between the numerical results for a chain of $L=40$ 
sites and the analytical results given by Eq.\eqref{wf}. The upper curves correspond to
$U_{IB} = U_{BB} + t_B$ and the lower ones correspond to $U_{IB} = U_{BB} - 2t_B$.
As expected, both results coincide in the strong-coupling limit.  
Note that the extra boson is displaced
over only a few lattice sites ($r \le 6$), so that this physics
can be realized in today's optical lattices which typically have
a linear size corresponding to 100 sites or so.

{\it General Case.} The bound states of ${\bar {\mathcal H}}$ can be found in any 
dimension by using the Green's function formalism \cite{economou2006}.  
We construct a basis of states that diagonalizes ${\bar {\mathcal H}}_{1{\bf K}}$ 
and we separate ${\bar {\mathcal H}}_{\bf K}$ into three terms, 
${\bar {\mathcal H}}_{\bf K}={\bar {\mathcal H}}_{0\bf K}+{\bar {\mathcal H}}^{+}_{\bf K}+
{\bar {\mathcal H}}^{-}_{\bf K}$, where
\begin{eqnarray}
{\bar {\mathcal H}}^{\pm}_{{\bf K}} &=&  \epsilon^{\pm} |{\bf K}, \psi^{\pm} \rangle \langle {\bf K},\psi^{\pm} |,
\end{eqnarray}
$\epsilon^{\pm} = (U_{\bf K} \pm \zeta_{\bf K})/2$, $\zeta_{\bf K}= \sqrt{U^2_{\bf K}+4\tau^2}$,
\begin{eqnarray}
| {\bf K},\psi^{\pm} \rangle &=& u_{\mp} |{\bf K}, {\bf 0}\rangle  \mp u_{\pm} |\psi_s \rangle, \;\;
| \psi_s \rangle = \frac{1}{\sqrt{Z}} \sum_{\langle {\bf 0}, {\bf r'} \rangle} |{\bf K}, {\bf r'} \rangle,
\nonumber
\end{eqnarray}
and $u_{\pm} = ( 1 \pm U_{\bf K}/\zeta_{\bf K})^{1/2}/\sqrt{2}$.
By introducing the Green operators 
\begin{eqnarray}
G^0_{\bf K}(z) &\equiv& \frac{1}{z- {\bar {\mathcal H}}_{0{\bf K}} },\;\;
G^{+}_{\bf K}(z) \equiv \frac{1}{z- {\bar {\mathcal H}}_{0{\bf K}} - {\bar {\mathcal H}}^{+}_{\bf K} },\nonumber\\
G_{\bf K}(z) &\equiv& \frac{1}{z- {\bar {\mathcal H}}_{{\bf K}} },
\end{eqnarray}
we obtain the $G^{+}_{\bf K}(z)$-operator from $G^0_{\bf K}(z)$ by expanding in ${\bar {\mathcal H}}^+_{{\bf K}}$:
\begin{eqnarray}
G^{+}_{\bf K} &=& G^0_{\bf K}  + G^0_{\bf K} | {\bf K}, \psi^+ \rangle  \frac{\epsilon^+}{1- \epsilon^+  
G^0_{\bf K} (\psi^+,\psi^+) }  \langle  {\bf K}, \psi^+  |  G^0_{\bf K}
\nonumber 
\end{eqnarray}
with $G^0_{\bf K} (\psi^+,\psi^+) = \langle  {\bf K}, \psi^+  |   G^0_{\bf K} | {\bf K}, \psi^+ \rangle$.
Similarly, by considering ${\bar {\mathcal H}}^{+}_{{\bf K}}$ as the unperturbed Hamiltonian 
and expanding in the perturbation  ${\bar {\mathcal H}}^-_{\bf K}$, we find
\begin{eqnarray}
G_{\bf K} &=& G^{+}_{\bf K}  + G^{+}_{\bf K} | {\bf K}, \psi^- \rangle  \frac{\epsilon^-}{1- \epsilon^-  
G^{+}_{\bf K} (\psi^-,\psi^-) }  \langle {\bf K}, \psi^-  |  G^{+}_{\bf K}, \nonumber \\
\label{gk}
\end{eqnarray}
with $G^{+}_{\bf K} (\psi^{-},\psi^{-}) = \langle  {\bf K}, \psi^{-}  |   G^{+}_{\bf K} | {\bf K}, \psi^{-} \rangle$.
The exact dispersion relation of the bound state is obtained from the poles of $G_{\bf K} ({\bf 0},{\bf 0}) 
\equiv \left\langle {\bf K}, {\bf 0} \left| G_{\bf K}\right| {\bf K},{\bf 0} \right\rangle$.

%%%%%%%%%%%%%%%%%%%%%%%%%%%
We consider $\epsilon^b_{\bf K}$, $\lambda_{\bf 0}$ and $m^*$ near the critical point, $U_{\bf 0}=U_c$.
%for the formation of a bound state. 
Identifying $\lambda_{\bf K}$ in $d$-dimensions from the asymptotic behavior
of the bound state wave function:
$\left\langle \psi^0_{\bf K} \right| \left. {\bf K}, {\bf r} \right\rangle \sim {e^{-r/\lambda_{\bf K}}}/{r^{(d-1)/2}}$
for ${\bf r}=(r,0,0,\cdots )$ and $r \gg 1$, we obtain 
%the $\lambda_{\bf K}$-dependence of the bound-state energy:  
\begin{equation}
	\epsilon^b_{\bf K} = -2t_B(n+1)\left[ d-1+\cosh (\lambda^{-1}_{\bf K})\right].
\end{equation}
Thus, $\lambda_{\bf 0} \propto \Delta^{-1/2}_b$ where $\Delta_b \equiv -2dt_B(n+1)-\epsilon^b_{\bf 0}$  
is the composite particle binding energy.  
From Eq. (\ref{eq:meffdef}),
we obtain the effective mass, $m^*$,
\begin{eqnarray}
\frac{m_I}{m^*}= \left. \frac{d \epsilon^b_{\bf K}}{d U_{\bf K}}\right|_{{\bf K}={\bf 0}}.
\end{eqnarray}
Near the critical point, the large polaron size washes out the dependence on specifics
that occur on the scale of the lattice constant.  Hence, we expect the binding energy to vanish
with the same scaling laws as the binding of a single impurity \cite{economou2006} 
%Since the microscopic details of ${\bar {\cal H}}$ near the impurity site become negligible for large 
%$\lambda_{\bf K}$, the bound state energy is expected to go to zero following the same behavior 
%as in the case of the conventional single impurity problem 
\footnote{
	The impurity only affects the diagonal energy of the site at the origin.
	%as ${\bar {\mathcal H}}_{1{\bf K}} =   u |{\bf K}, {\bf 0} \rangle \langle {\bf K}, {\bf 0}|$.
}:
\begin{eqnarray}
\Delta_b ({\bf 0})  &\propto& (U_{\bf K}-U_c)^{2} \;\; {\rm for} \;\;d=1 \;{\rm and}\; 3, \nonumber\\ 
\Delta_b ({\bf 0})  &\propto& \exp
	\left[	\frac{C}{U_{\bf 0}-U_c}
	\right], \;\; {\rm for}\;\; d=2, 
\end{eqnarray}
where the constant $C$ depends on microscopic details of ${\bar {\cal H}}$.
These equations lead to the relation between $m^*$ and $\lambda_{\bf 0}$:
\begin{eqnarray}
\frac{m^*}{m_I}   &\propto& \lambda_{\bf 0}  \;\; {\rm for} \;\;d=1 \;{\rm and}\; 3, \nonumber\\ 
\frac{m^*}{m_I} &\propto& \lambda_{\bf 0}^2 (\ln{\lambda_{\bf 0}})^2  m_I \;\; {\rm for}\;\; d=2. \nonumber
\end{eqnarray}

%%%%%%%%%%%%%%%%%%%%%%%%%
 \begin{figure}[htb]
 \begin{center}
 \includegraphics[trim= 0 0 0 0,angle=270,width=8cm,clip]{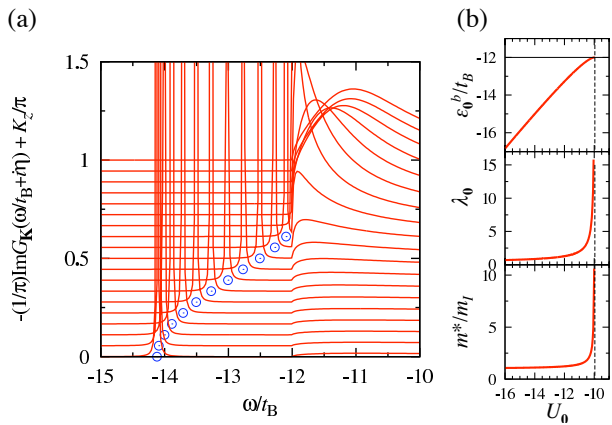}
\caption{
(a) $K_z$ dependence of spectral weight for the simple cubic lattice
($t_B=t_I=1$, $n=1$, $K_x=K_y=0$, and $U_{BB}-U_{IB}=7$).
Each line is shifted by $K_z/\pi$ for visualization. 
We use $\eta=2^{-10}$ as an infinitesimal constant.
(b) $U_{\bf 0}$ dependences of the bound state energy $\epsilon^b_{\bf 0}$ (top),
linear size of the polaron  $\lambda_{{\bf 0}}$ (middle), 
and ratio of effective and bare masses $m^*/m_I$ (bottom) for $n=$1.
The dashed line indicates $U_{\bf 0}= U_c$ .}
 \label{3dDOS}
 \end{center}
 \end{figure}
%For concreteness we consider now the case of a simple cubic lattice ($d=3$).
For the case of a simple cubic lattice in $d$=3,
Fig. \ref{3dDOS} shows the spectral density obtained from Eq.\eqref{gk}.
The bound states for different values of ${\bf K}$ that appear below the bottom 
of continuum spectrum ($U_{\bf K}<U_c$)
form the polaron band. $U_c$  is obtained by finding a root of 
$1- \epsilon^-  G^{+}_{\bf K} (\psi^-,\psi^-) =0$ at $z=-2dt_B(n+1)$ as a function of $n$.
%By computing $U_c(n)$ for the simple cubic lattice we verified that
%it converges to the critical value of conventional impurity problem 
%(uniform hopping amplitude) for $n \to \infty$.
%This is the expected result because the relative contribution of 
%$\tau$, $\tau/\{(n+1)|t_{B}|\} \propto 1/n$, becomes negligibly small 
%for $n \to \infty$.
Figure \ref{3dDOS}(b) shows $U_{\bf 0}$ dependences of $\epsilon^b_{\bf 0}$, $\lambda_{\bf 0}$ 
and $m^*/m_I$ for $n=1$.  Note that the effective mass $m^*$ 
and the polaron size increase sharply with decreasing $U_c-U_{\bf 0}$.
By superimposing a linear or harmonic impurity-specific potential and observing the subsequent acceleration or oscillation,
cold atom experiments can measure $m^*$ directly.

The divergence of the effective mass for $U_{\bf 0} \to U_c$ is cut off at a finite value 
$m^*_{\rm max} = U_{IB}/(2nt_I t_B)$ when the next order correction in $t_{\nu}/U_{IB}$ 
is included implying that $m^*$ can be tuned over a large spectrum of values 
ranging from $m_I= 1/2t_I$ to $ U_{IB} m_I/(2n t_B)$. On the other hand,  the
polaron size can be tuned from zero to infinity by changing the effective potential 
$U_{\bf 0}$ or, equivalently, the difference between $U_{IB}$ and $U_{BB}$. Such
dependence of $m^*$ on the particle-hole pair size is not shared by the usual 
two-particle bound states such as condensed matter excitons. In those cases the 
mass of the bound state is well approximated by the sum of the masses. 
The Mott polaron mass increases with size because, to linear order in $t_{I}$,
the impurity can only move
when the particle and hole share the same site (see Fig.\ref{lowenstates}(b)). 
This strongly correlated mechanism leads to a highly tunable $m^*$.

The control offered by cold atom technology over polaron size and mass hints at the intriguing
prospect of studying the quantum phase transition between a ``confined" gas of 
boson-impurity polarons and two ``deconfined'' gases of impurities and bosons 
in the next Hubbard band. The driving parameter of this transition is the difference 
$U_{IB}-U_{BB}$. By reducing the ratio $U_{IB}/t_B$, while keeping  $U_{\bf 0}$ fixed,
it would be possible to vary the polaron mass near the transition. 
The competition between potential and kinetic energies 
could also lead to an intermediate  crystallization of polarons  in  
the large $m^*$ regime.

We are grateful to S. A. Trugman for many useful discussions. 
Work at the LANL was performed under the auspices of the
U.S.\ DOE contract No.~DE-AC52-06NA25396 through the LDRD program. A.E.F.
thanks NSF for support through Grant No. DMR-0955707.

%\newpage %Just because of unusual number of tables stacked at end
\bibliography{optpola}% Produces the bibliography via BibTeX.

\end{document}